\documentclass[sigconf]{acmart}

\usepackage{booktabs} 
\usepackage{amssymb}
\usepackage{amsmath}
\usepackage[ruled,linesnumbered]{algorithm2e}

\DeclareMathAlphabet{\mathcal}{OMS}{ntxsy}{m}{n}   

\usepackage{caption}
\usepackage{subcaption}
\usepackage{float}
\newcommand{\usr}{u}
\newcommand{\Usr}{\mathcal{U}}
\newcommand{\Edg}{\mathcal{E}}
\newcommand{\loc}{\ell}
\newcommand{\Loc}{\mathcal{L}}
\newcommand{\w}{w}
\newcommand{\cicnt}[1]{\vert{\it ci}(#1)\vert}

\newcommand{\umat}{X}
\newcommand{\rumat}{\overline X}
\newcommand{\ulmat}{Y}
\newcommand{\rulmat}{\overline Y}
\newcommand{\lumat}{Y^{T}}
\newcommand{\rlumat}{\overline{Y^{T}}}

\newcommand{\ls}[1]{\kappa(#1)}
\newcommand{\ui}[1]{\eta(#1)}
\newcommand{\vecls}{\kappa}
\newcommand{\vecui}{\eta}

\newlength{\thinline}
\newlength{\thickline}
\copyrightyear{2017}
\acmYear{2017}
\setcopyright{acmcopyright}
\acmConference{HT'17}{}{July 4-7, 2017, Prague, Czech Republic.}
\acmPrice{15.00}
\acmDOI{http://dx.doi.org/10.1145/3078714.3078729}
\acmISBN{978-1-4503-4708-2/17/07}

\begin{document}
\title{Quantifying Location Sociality}

\author{Jun Pang}
\affiliation{%
  \institution{FSTC \& SnT\\ University of Luxembourg}
}
\email{jun.pang@uni.lu}

\author{Yang Zhang}
\affiliation{%
  \institution{
  CISPA, Saarland University\\
Saarland Informatics Campus
}
}
\email{yang.zhang@cispa.saarland}

\begin{abstract}
The emergence of location-based social networks provides 
an unprecedented chance to study the interaction 
between human mobility and social relations.
This work is a step towards 
quantifying whether a location is suitable for conducting social activities,
and the notion is named location sociality.
Being able to quantify location sociality creates practical opportunities
such as urban planning and location recommendation.
To quantify a location's sociality,
we propose a mixture model of HITS and PageRank
on a heterogeneous network linking users and locations.
By exploiting millions of check-in data generated 
by Instagram users in New York and Los Angeles,
we investigate the relation 
between location sociality and several location properties, 
including location categories, rating and popularity.
We further perform two case studies, i.e., friendship prediction and location recommendation,
experimental results demonstrate the usefulness of our quantification.
\end{abstract}

%
%



\maketitle

\section{Introduction}
\label{sec:intro}

Online social networks (OSNs) have been the most successful web applications 
during the past decade.
Leading companies, including Facebook,\footnote{\url{https://www.facebook.com/}} 
Twitter\footnote{\url{https://twitter.com/}} 
and Instagram,\footnote{\url{https://www.instagram.com/}}
have gained a large number of users.
More recently, 
with the development of positioning technology on mobile devices,
OSNs have been extended to geographical space.
Nowadays, it is quite common for OSN users to share their geographical locations, 
i.e., check-ins.
Moreover, a special type of OSNs dedicated to location sharing are created,
namely location-based social networks (LBSNs).
Foursquare and Yelp are two representative companies.
With the emergence of LBSNs,
a large quantity of data concerning human mobility become available.
This gives us an unprecedented opportunity to understand human mobility
and moreover to study the interaction 
between social relations and mobility.
Some previous works have been focused on inferring social relationships from mobility,
such as~\cite{SNM11,PSL13,WLL14,HYL15,ZP15},
others exploit users' social information to predict their future locations,
such as~\cite{BSM10,CML11}.
More recently, researchers propose new understandings of locations 
by using user generated data
such as happiness~\cite{QSA14} and walkability~\cite{QASD15}.

Location has been recognized as an important factor for social activities back in 1950s.
In his seminal work~\cite{G59}, Erving Goffman 
described social interactions as a series of performance given by social actors,
and  physical setting, i.e., location, is an important aspect of a social actor's performance.
In~\cite{G59}, Goffman stated that ``\emph{A setting tends to stay put, geographically speaking, 
so that those who would use a particular setting as a part of their performance 
cannot begin their act until they have brought themselves to the appropriate place}''.
Based on Goffmann's study, Milligan~\cite{M98} further proposed that
``\emph{physical sites (however defined by the participants) 
become the stages for social interaction, 
stages that are both physically and socially constructed}''.
She explained that not only being physically constructed
(by architects, facility managers, property owners and others),
a location will also be socially constructed by people who conduct social interactions there.
Following this theory, we argue that social construction 
will make some locations more suitable for social activities than others.

In the current work, we aim to quantify 
whether a location is a suitable for conducting social activities.
The notion we quantify is named \emph{location sociality}.
We define a location's sociality as \emph{the degree 
to which individuals tend to conduct social activities at that location}.
A location is considered social if friends frequently visit, 
especially for the purpose of socializing or recreation, and vice versa.
Studying location sociality could advance the boundary 
of our understanding on the interaction between social relations and mobility.
It can also help us to solve challenging problems such as urban planning and traffic control.
In practice, location sociality can be also used to build appealing applications
such as location recommendation.

\medskip
\noindent\textbf{Contributions.}
In the current work, we make the following contributions:
\begin{itemize}
 \item  We propose a framework to quantify location sociality (Section~\ref{sec:framework}).
 Our framework is based on the assumption that
 a location's sociality and its visitors' social influence
 are mutually reinforced.
 To model this assumption, we construct a heterogeneous network 
 consisting of users (in a social network) and locations (a user-location network).
 Then, we propose a mixture model of HITS~\cite{K99} and PageRank 
 to quantify location sociality on this heterogeneous network.
 
 \item Following our solution, we exploit millions of check-in data from Instagram
 in New York and Los Angeles to quantify location sociality (Section~\ref{sec:exp}).
 We then study the relation between location sociality and several location properties
 including location categories, rating and popularity.
 Our discoveries include:
 certain types of locations (music venues and nightclubs) are more social than others;
 location sociality shares a positive relation with location rating given by users;
 social locations distribute more uniformly w.r.t.\ geographical space than popular locations.
 
 \item To demonstrate the usefulness of our quantification of location sociality,
 we perform a case study on friendship prediction in Section~\ref{sec:linkpre}.
 We extract two users' common locations 
 and define features based on these common locations' sociality
 for machine learning classification.
 Experimental results show that with very simple location sociality features,
 we are able to achieve a strong prediction.
 Moreover, adding location sociality into a state-of-the-art prediction model 
 achieves a 5\% performance gain.
 
 \item We perform another case study 
 on using location sociality for location recommendation in Section~\ref{sec:locrecomm}.
 We integrate our quantification into a random walk with restart framework.
 Experimental results show that 
 the recommender based on location sociality 
 achieves a better recommendation performance (at least 5\%)
 than the baseline recommender that does not consider location sociality.
\end{itemize}

We discuss some implications and limitations of the current work in Section~\ref{sec:implimit}.
Related works are discussed in Section~\ref{sec:rel} and
Section~\ref{sec:conclu} concludes the paper.

\section{Proposed Solution}
\label{sec:framework}

In this section, we first discuss the intuition 
of our solution on quantifying location sociality (Section~\ref{ssec:intuition}),
then we formally describe the solution (Section~\ref{ssec:framework}).

\subsection{Intuition}
\label{ssec:intuition}

Our intuition on quantifying location sociality in this paper
is based on the assumption
that a location's sociality and its visitors' social influence are mutually reinforced.
To explain this intuition, we start by addressing socially influential users.
In the society, if a person is considered socially influential, 
he must visit different social places frequently
to organize or participate in different social activities and events.
On the other hand, if a location is frequently visited by influential users,
then it must be suitable for conducting social activities, i.e., it is a social place.
Following this, we establish a mutual reinforcement relation 
between user influence and location sociality,
i.e., more social a location is, more socially influential users visit it, and vice versa.
In addition to visiting many social places,
an influential user should also occupy an important position in the social network, 
e.g., he should have many friends who are also socially influential.
Following the above discussion, our intuition on quantifying location sociality can be summarized 
as the following two assumptions. 

\smallskip
\noindent\emph{Assumption 1.} Location sociality and users' social influence are mutually reinforced.

\smallskip
\noindent\emph{Assumption 2.} Users' social influence can be quantified from the social network.

\smallskip
The first intuition can be naturally formulated into a HITS-style framework~\cite{K99}.
For the second intuition, we apply PageRank on the social graph
to quantify each user's influence.
In the rest of the paper, we use location sociality and sociality interchangeably.

\subsection{Our framework}
\label{ssec:framework}

We start by modeling users, locations and their relationships into two types of networks
including social network and user-location network.

\smallskip
\noindent\textbf{Social network.} A \emph{social network}, denoted as $G_{\Usr} = (\Usr, \Edg_{\Usr})$,
is an unweighted graph with nodes in set $\Usr$ representing all users.
$\Edg_{\Usr} \subseteq \Usr\times \Usr$ is a symmetric relation containing the edges in $G_{\Usr}$.
If $\usr_i$ and $\usr_j$ are friends, then $(\usr_i, \usr_j)\in \Edg_{\Usr}$
and $(\usr_j, \usr_i)\in \Edg_{\Usr}$.
We use matrix $\umat$ to represent $G_{\Usr}$
where $\umat_{i,j} = 1$ if $(\usr_i, \usr_j)\in \Edg_{\Usr}$
and $\umat_{i,j} = 0$ otherwise.
It is easy to see that $\umat$ is symmetric.
We further use $\rumat$ to denote the column stochastic matrix of $\umat$
where $\rumat_{i, j} = \frac{\umat_{i, j}}{\sum_k \umat_{k, j}}$.

\begin{figure}[!t]
  \centering
  \includegraphics[width=1.0\columnwidth]{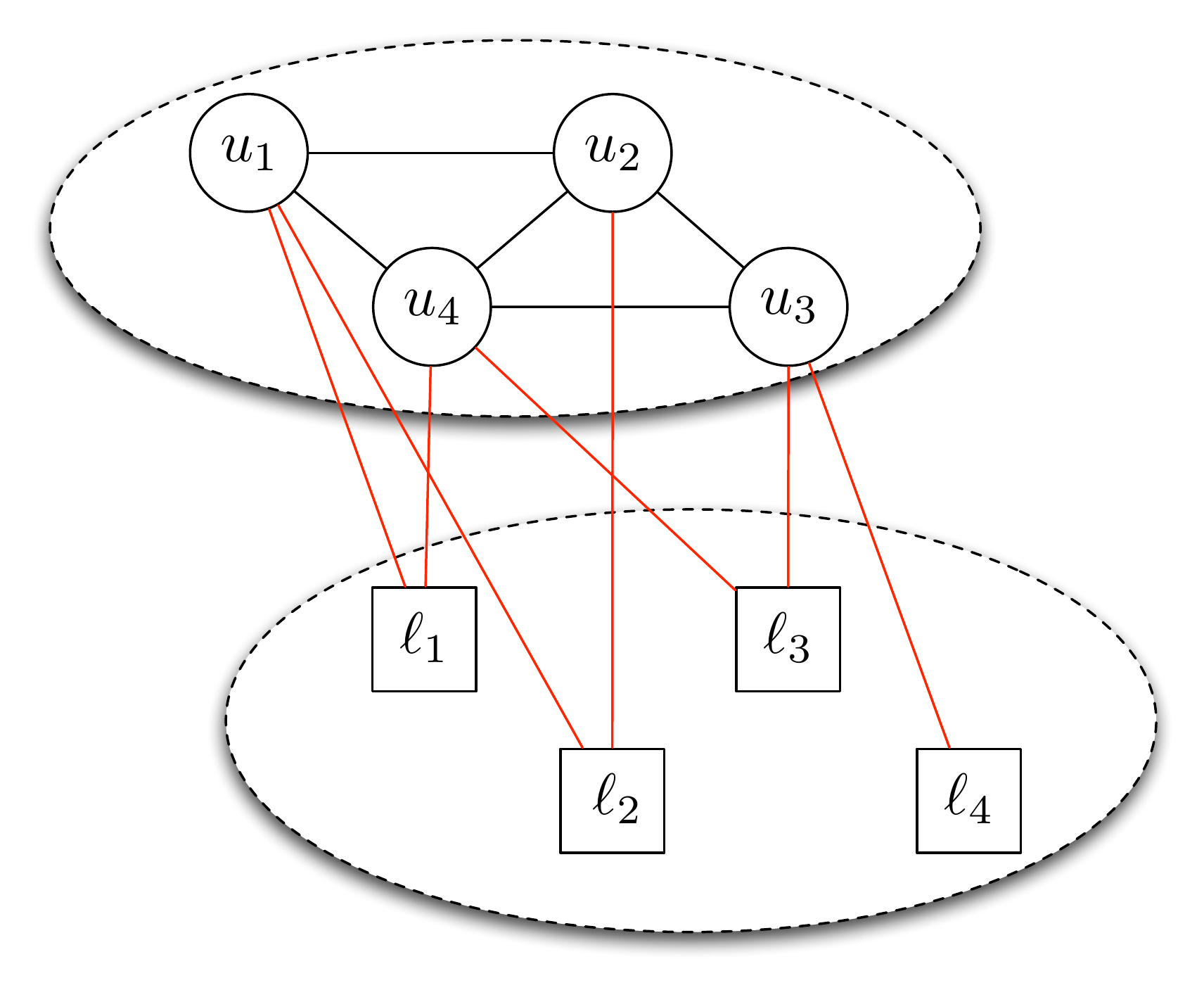}
  \caption{A model of the network. \label{fig:model}}
\end{figure}

\smallskip
\noindent\textbf{User-location network.}
A \emph{user-location network}, denoted as $G_{\Usr, \Loc}$ $= (\Usr, \Loc, \Edg_{\Usr, \Loc})$,
is a weighted bipartite graph.
$\Edg_{\Usr, \Loc} \subseteq \Usr\times \Loc$ consists of the edges in $G_{\Usr, \Loc}$.
Each edge $(\usr_i,\loc_j)\in \Edg_{\Usr, \Loc}$, also written as $e^{\Usr, \Loc}_{i, j}$,
is associated with a weight $\w^{\Usr, \Loc}_{i, j}$ defined as the number of times
that the user $\usr_i$ has visited (checked in) the location $\loc_j$ (denoted by $\cicnt{\usr_i, \loc_j}$).
We use matrix $\ulmat$ to represent $G_{\Usr, \Loc}$
with $\ulmat_{i, j} = \w^{\Usr, \Loc}_{i, j}$.
The transpose of $\ulmat$ is further denoted by $\lumat$.
In the end, we use $\rulmat$ and $\rlumat$ to denote the column stochastic matrices of $\ulmat$ and $\lumat$, respectively.

Figure~\ref{fig:model} shows an example of the heterogeneous graph.
Within our framework two sets of values, locations' sociality and users' social influence, can be obtained.
Each location $\loc$'s sociality is defined as $\ls{\loc}$
and $\ui{\usr}$ for each user's social influence.
Following the intuition in Section~\ref{ssec:intuition},
our model is formulated into the following equations:

\begin{minipage}{0.8\linewidth}
\begin{equation}\label{equ:pru}
 \ui{\usr_i} = \sum_j\!{\rumat}_{i, j}\cdot \ui{\usr_j}
\end{equation}
\end{minipage}

\begin{minipage}{0.8\linewidth}
\begin{equation}\label{equ:hitsu}
 \ui{\usr_i} = \sum_j\!{\rulmat}_{i, j} \cdot \ls{\loc_j}
\end{equation}
\end{minipage}

\begin{minipage}{0.8\linewidth}
\begin{equation}\label{equ:hitsl}
 \ls{\loc_j} = \sum_i\!{\rlumat}_{j, i} \cdot \ui{\usr_i}
\end{equation}
\end{minipage}

Equations~\ref{equ:pru} is the PageRank implementation 
for quantifying users' social influence from $G_{\Usr}$.
Equations~\ref{equ:hitsu} and~\ref{equ:hitsl} are an instance of the HITS framework
which establishes the mutual reinforcement relationship between locations and users.
We then linearly combine the above equations as

\begin{minipage}{0.8\linewidth}
\begin{equation}
 \ui{\usr_i} = \alpha\cdot \sum_j {\rumat}_{i, j}\cdot \ui{\usr_j} + 
 (1-\alpha)\cdot \sum_j {\rulmat}_{i, j}\cdot \ls{\loc_j}
\end{equation}
\end{minipage}

\begin{minipage}{0.8\linewidth}
\begin{equation}\label{equ:combls}
 \ls{\loc_j} = \sum_i {\rlumat}_{j, i}\cdot \ui{\usr_i}
\end{equation}
\end{minipage}

\smallskip\noindent
where $\alpha$ specifies the contributions of each component to users' social influence.
In our experiments, $\alpha$ is set to 0.5
which indicates the social network structure and user mobility are equally important 
on quantifying users' social influence.
Note that $\alpha=0.5$ is a typical setting in many fields such as~\cite{WYX07}
where the authors aim to discover salient sentences for document summarization.

We further use two vectors $\vecui$ and $\vecls$ 
to denote users' social influence and locations' sociality.
Then the above equations can be written into the following matrix form.

\begin{minipage}{0.8\linewidth}
\begin{equation}\label{equ:matu}
 \vecui = \alpha \cdot \rumat \cdot \vecui + (1-\alpha) \cdot \rulmat \cdot \vecls
\end{equation}
\end{minipage}

\begin{minipage}{0.8\linewidth}
\begin{equation}\label{equ:matl}
 \vecls =  \rlumat \cdot \vecui
\end{equation}
\end{minipage}

\smallskip\noindent
Equations~\ref{equ:matu} and \ref{equ:matl} can be computed through an iterative updating process.
We set all locations' (users') initial sociality (social influence) 
to be $\frac{1}{\vert \Loc\vert}$ ($\frac{1}{\vert \Usr\vert}$).
According to our experiments, the computation stops after around 10 iterations,
when the maximal difference between $\vecls$s of two consecutive iterations is less than 0.00001.

\section{Experiments}
\label{sec:exp}

In this section, we first introduce the dataset used for our experiments.
Then, we present the results of our quantification:
we start by discussing the top social locations and location categories;
then we focus on the relation between location sociality and location rating;
in the end, the correlation between location sociality and popularity is discussed.

\subsection{Dataset description}
Instagram is a photo-sharing social network
with a fast growing user number. 
By now, it has 400M monthly active users and with 75M photos published everyday.
Similar to other social network services such as Facebook and Twitter, 
Instagram allows users to share their locations when publishing photos.
Moreover, unlike Twitter where only a small amount of tweets are geo-tagged,
the authors of~\cite{MHK14} have shown that Instagram users 
are much more willing to share their locations (31 times more than Twitter users),
which makes Instagram a suitable platform to study 
the interaction between mobility and social relations.

\begin{figure}[!t]
  \centering
  \includegraphics[width=0.85\columnwidth]{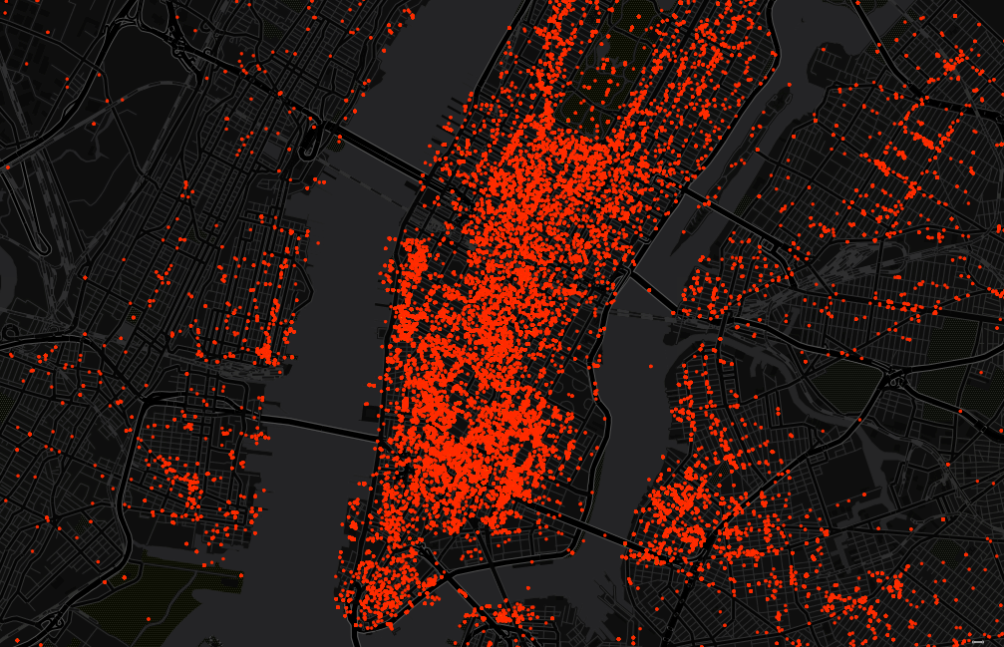}
  \caption{Check-ins in New York.\label{fig:nycheckin}}
\end{figure}

We collect the geo-tagged photos, i.e., check-ins, in New York and Los Angeles
from Instagram through its public API\footnote{\footnotesize\url{https://www.instagram.com/developer/}}.
Since locations' category information is an important aspect of our analysis,
and fortunately the API of Instagram is linked with the API of Foursquare, 
a leading location-based social network with resourceful information about each place,
thus we exploit the following methodology to collect our data.
We first resort to Foursquare
to extract all location ids within each city,
meanwhile we collect each location's category information together with its rating (number of tips and number of likes).
Then for each Foursquare's location id,
we query Instagram's API to get its corresponding location id in Instagram.
After this,
we query each location's recent check-ins in Instagram several times a day
from August 1st, 2015 until March 15th, 2016. 
In the end, more than 6M check-ins have been collected in New York
and 4.7M in Los Angeles\footnote{It is worth 
noticing that the authors of~\cite{MHNW15} 
has applied a similar methodology.}.
To resolve the data sparseness issue, 
we focus on users with at least 20 check-ins (considered as active users)
and locations with at least 10 check-ins.
Figure~\ref{fig:nycheckin} depicts a sample check-in distribution in New York.
Since Foursquare organizes location categories into a tree structure\footnote{\footnotesize
\url{https://developer.foursquare.com/categorytree}},
we take its second level categories to label each location.

\begin{table}[!t]
 \centering
 \begin{tabular}{ l | r | r}
  \toprule[\thickline]
 & ~~~ New York & ~~~ Los Angeles \\
  \midrule[\thinline]
 \# check-ins & 6,181,169 & 4,705,079 \\
 \hline
 \# active users & 12,280 & 8,643 \\
 \hline
 \# edges (active users) & 74,230 & 44,994\\
 \hline
 \# locations & 8,683 & 6,908\\
 \bottomrule[\thickline]
 \end{tabular}
 \captionof{table}{Dataset summary.~\label{table:dataset}}
\end{table}

To obtain users' social networks, 
we exploit Instagram's API
to query each active user's follower/followee list\footnote{Since Instagram's API only provides one page with 50 follower/followees per query,
we perform multiple queries until all follower/followees of each user are obtained.}.
We consider two users as friends if they mutually follow each other in Instagram.
To further guarantee that users we have collected are not celebrities or business accounts,
we filter out the top 5\% of users with most followers.
Also,  only the relations among active users (users with at least 20 check-ins) are kept.
In the end, the social network contains 74,230 edges for New York and 44,994 edges for Los Angeles.
Table~\ref{table:dataset} summarizes the dataset.
For the sake of experimental result reproducibility,
our dataset is available upon request.

\subsection{Location sociality vs.\ location category}
\begin{figure}[!t]
\centering
\begin{minipage}{0.85\columnwidth}
  \includegraphics[width=\columnwidth]{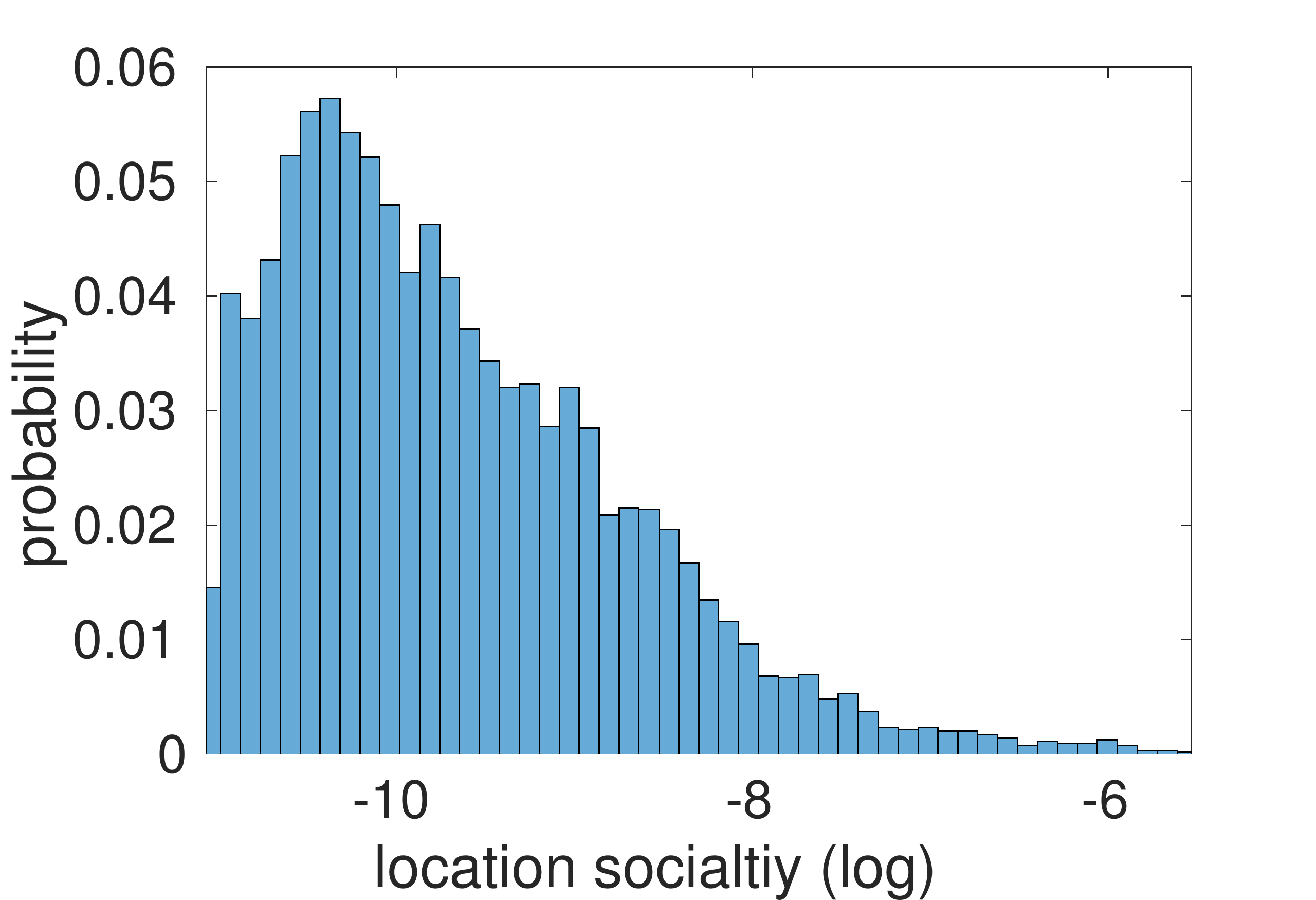}
  \subcaption{New York\label{fig:ny_histls}}
\end{minipage}
\begin{minipage}{0.85\columnwidth}
  \includegraphics[width=\columnwidth]{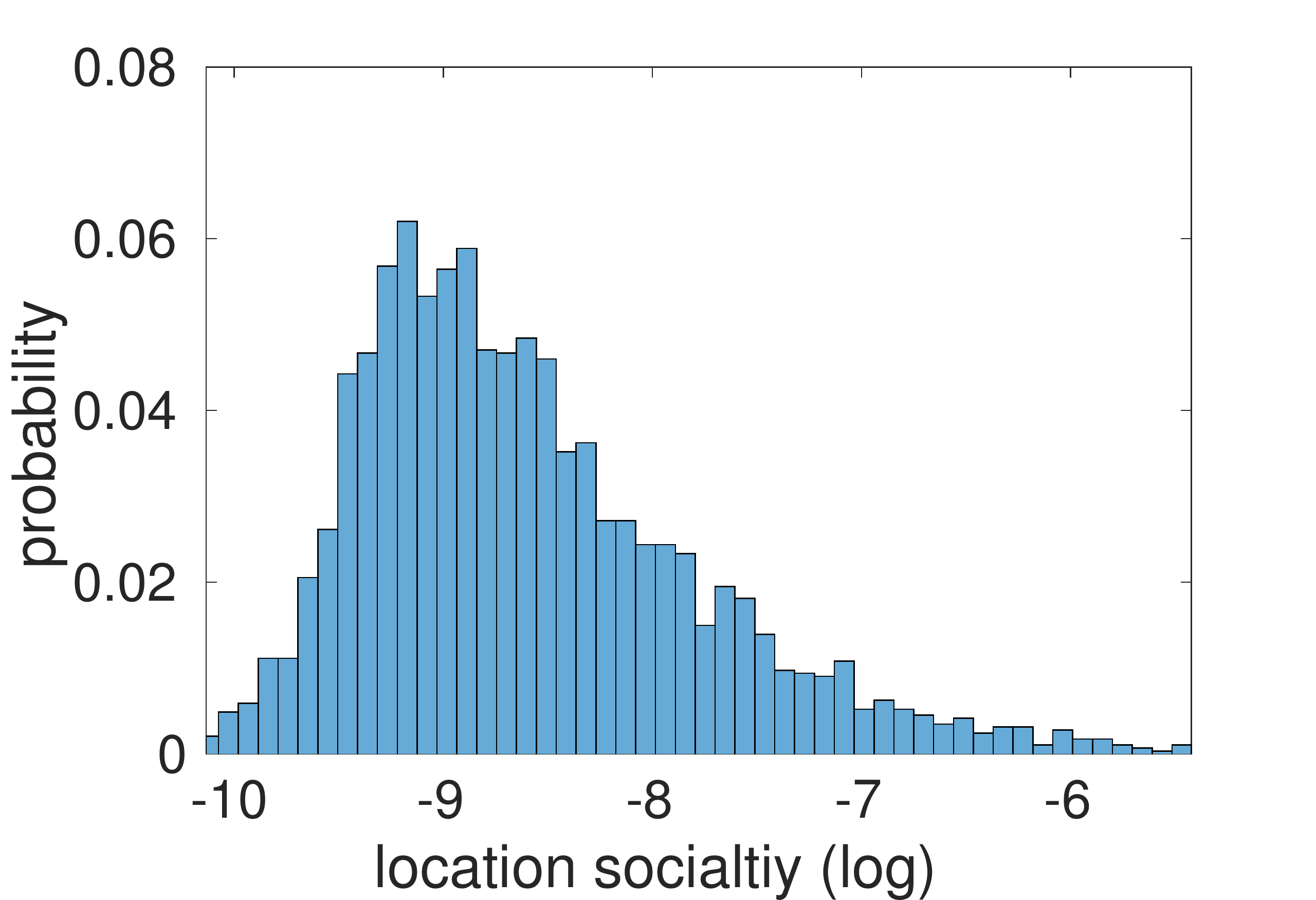}
  \subcaption{ Los Angeles\label{fig:la_histls}}
\end{minipage}
\caption{Distributions of log-transformed location sociality.\label{fig:histls}}
\end{figure} 

Figure~\ref{fig:histls} 
depicts the log transformed distributions of location sociality,
both of which 
indicate that most locations have a middle value of sociality
while only a few locations are very social or unsocial.
This is different from other location measurement,
for instance, the number of mobility transitions 
from or to each location follows a power law distribution~\cite{NSLM15}.

Table~\ref{table:toploc} presents the top five locations
with highest and lowest sociality.
For New York, Webster Hall (music venue) is the most social place
followed by Madison Square Park (park).
On the other hand, the least social place is one Staples store (convenience store) in midtown.
For Los Angeles, The Fonda Theatre (concert hall) has the highest sociality.
Meanwhile, one Panda Express (fast food restaurant)
is the least social place.
From Table~\ref{table:toploc},
we can see a clear distinction between social and unsocial places w.r.t.\ their categories.
Next, we take a deeper look at the relation between sociality and location category.

\begin{table}[!t]
\centering
 \begin{tabular}{c|c}
\toprule[\thickline]
 \multicolumn{2}{c}{\textbf{New York}}\\
 \midrule[\thinline]
 Top Social & Top Unsocial \\
\midrule[\thinline]
 Webster Hall & Staples\\
 Madison Square Park &  17 Frost Gallery\\
 Rockwood Music Hall & China Institute\\
 Washington Square Park & Manhattan Theatre Club \\
 Baby's All Right & El Rey Del Taco II\\
\bottomrule[\thickline]
\toprule[\thickline]
 \multicolumn{2}{c}{\textbf{Los Angeles}}\\
 \midrule[\thinline]
 Top Social & Top Unsocial \\
\midrule[\thinline]
 The Fonda Theatre & Panda Express\\
 Avalon Hollywood & Gap\\
 The Echo & Ebar\\
 Hermosa Beach Pier & Palms Super Market\\
 Exchange LA & 7-Eleven\\
\bottomrule[\thickline]
\end{tabular}
\caption{The most and least social locations in New York and Los Angeles.\label{table:toploc}}
\end{table}

Table~\ref{table:topcat} lists the top five location categories 
with the highest and lowest average location sociality.
Nightclub and music venue are in the top 3 in both cities.
On the other hand, convenience store seems to be less attractive to friends.
Besides, we also observe some interesting difference between the two cities.
For example, beach is the No.\! 5 social choice for people living in Los Angeles
while it is not New Yorkers' choice since there are no beaches in Manhattan.

\begin{table}[!t]
 \centering
 \begin{tabular}{c|c}
\toprule[\thickline]
 \multicolumn{2}{c}{\textbf{New York}}\\
 \midrule[\thinline]
 Social Categories & Unsocial Categories\\
\midrule[\thinline]
 Music Venue & Laundry Service \\
 Nightclub &  Convenience Store \\
 Harbor &  Post Office \\
 Museum &  Pharmacy \\
 Park &  ~Fast Food Restaurant \\
\bottomrule[\thickline]
\toprule[\thickline]
 \multicolumn{2}{c}{\textbf{Los Angeles}}\\
 \midrule[\thinline]
 Social Categories & Unsocial Categories\\
\midrule[\thinline]
 Concert Hall &  Convenience Store\\
 Nightclub &  Vintage Store \\
 Music Venue &  ~Fast Food Restaurant~ \\
 Mall &  Pet Service \\
 Beach &  Automotive Shop \\
\bottomrule[\thickline]
 \end{tabular}
 \caption{Top 5 location categories with highest and lowest average location sociality 
 in New York and Los Angeles.\label{table:topcat}}
\end{table}

As music venue and nightclub have high rankings in both cities,
we further list the top 5 music venues and nightclubs
in Table~\ref{table:topcat2}.
Although the ranking of music venues and nightclubs are rather subjective,
we have checked several blogs and articles (listed in the additional material)
and most of our top-social nightclubs and music venues have received positive reviews 
and been recommended by these blogs and articles.
We conclude that a location's sociality is related to its category.
\begin{table}[!t]
 \centering
 \begin{tabular}{c|c}
\toprule[\thickline]
 \multicolumn{2}{c}{\textbf{New York}}\\
 \midrule[\thinline]
 Social Music Venues & Social Nightclubs\\
\midrule[\thinline]
 Webster Hall & Stage 48 \\
 Rockwood Music Hall &  Marquee \\
 Baby's All Right & Pacha NYC  \\
 Bowery Ballroom & 1 OAK \\
 Music Hall of Williamsburg~ & ~VIP Room NYC \\
\bottomrule[\thickline]
\toprule[\thickline]
 \multicolumn{2}{c}{\textbf{Los Angeles}}\\
 \midrule[\thinline]
 Social Music Venus & Social Nightclubs\\
\midrule[\thinline]
 Avalon Hollywood &  Exchange LA\\
 The Echo &  OHM Nightclub \\
 The Roxy &  Sound Nightclub \\
 The Troubadour &  Club Los Globos \\
 The Hollywood Bowl~ &  ~Create Nightclubs \\
\bottomrule[\thickline]
 \end{tabular}
 \caption{Top 5 music venues and nightclubs with highest location sociality 
 in New York and Los Angeles.\label{table:topcat2}}
\end{table}
Normally, location category itself is not sufficient to judge whether a location is social or not.
Next, we study other properties of locations
and their relationship with location sociality.

\subsection{Location sociality vs.\ rating, tips and likes}
For each location, Foursquare provides us with not only its category information,
but also other properties including rating\footnote{In Foursquare, rating is in the range from 1 to 10.},
number of tips and number of likes generated by Foursquare users.
Next, we study whether it is possible 
to use these properties to explain location sociality.
To proceed, we build a linear regression model 
with rating, number of tips and number of likes as explanatory variables
while location sociality as the dependent variable.
By fitting the model with ordinary least square method,
we obtain a coefficient of determination ($R^2$) of 0.192 in New York
and 0.280 in Los Angeles,
meaning that 19.2\% (28.0\%) of the variability of location sociality in New York (Los Angeles)
can be explained by these properties.
By checking the parameters of our linear model,
we discover that the major predictive power is driven by location rating.

\begin{figure}[!t]
\centering
\begin{minipage}{0.85\columnwidth}
  \centering
  \includegraphics[width=\columnwidth]{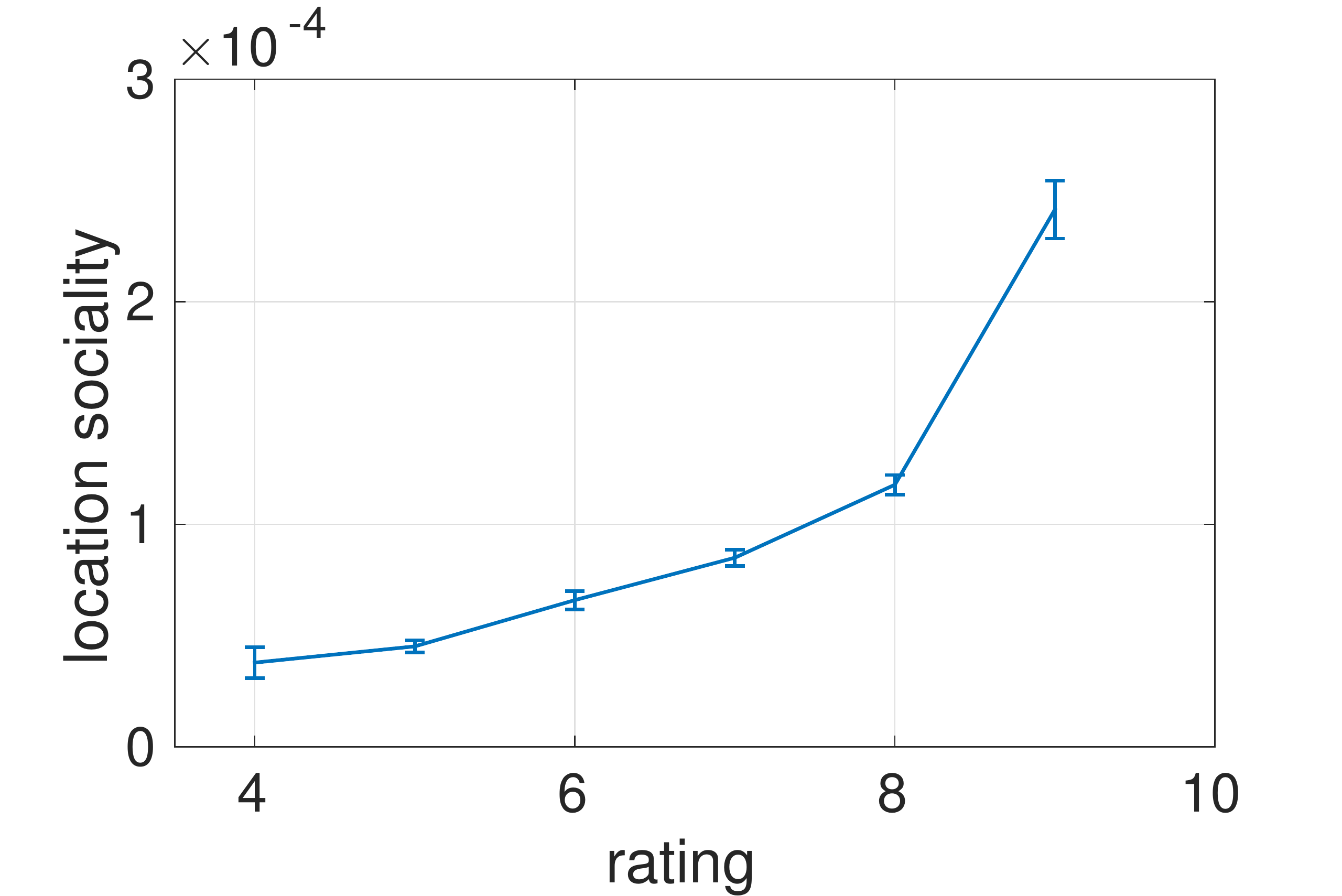}
  \subcaption{New York.\label{fig:ny_lsvsrating}}
\end{minipage}
\begin{minipage}{0.85\columnwidth}
  \centering
  \includegraphics[width=\columnwidth]{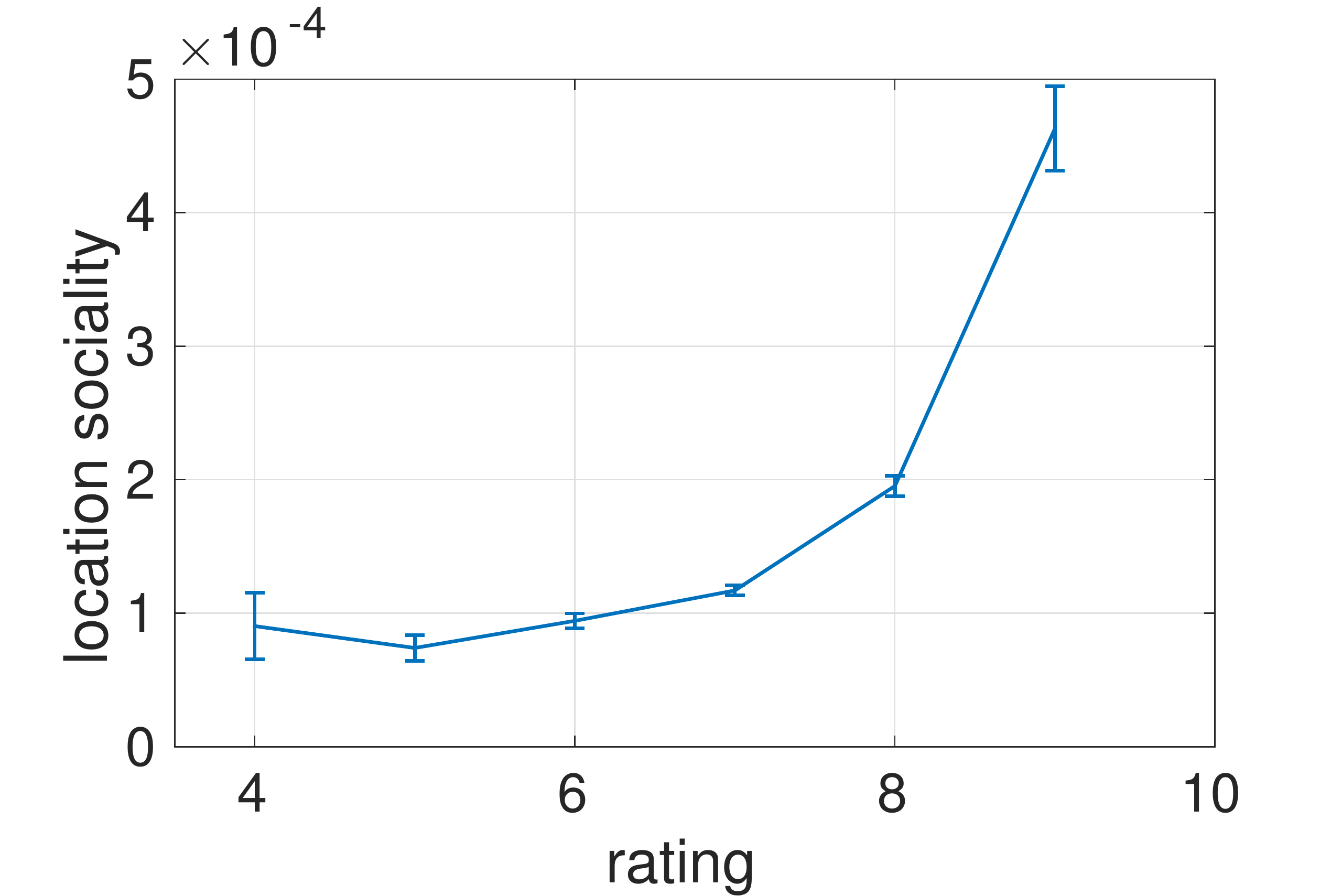}
  \subcaption{Los Angeles.\label{fig:la_lsvsrating}}
\end{minipage}
\caption{Location sociality vs.\ location rating.\label{fig:lsvsrating}}
\end{figure}

We further plot the average location sociality as a function of rating
in Figure~\ref{fig:lsvsrating}:
the two variables share a positive relation.
Especially when location rating is high ($\geq 8$),
location sociality increases sharply for both cities.
This indicates that social places are assigned with high ratings by users.

\subsection{Location sociality vs.\ location popularity}
\label{sec:lsvsle}
A social location is often popular in the sense that it attracts many people.
On the other hand, to conduct social activities, 
everyone has his own preference on choosing locations.
For example, one may prefer to go to a bar near his home, which might not be well-known at the city level.
The relationship between a location's sociality and its popularity is worth investigation:
a location's sociality should be correlated with its popularity,
while the two notions should exhibit difference.

By far, the most common notion for quantifying a location's popularity is location entropy~\cite{CTHKS10},
it is formally defined as
\[
 {\it le}(\loc) = -\sum\frac{\cicnt{\usr, \loc}}{\cicnt{\loc}}\log\frac{\cicnt{\usr, \loc}}{\cicnt{\loc}},
\]
where $\cicnt{\usr, \loc}$ is user $\usr$'s number of check-ins at location $\loc$ (Section~\ref{sec:framework})
and $\cicnt{\loc}$ is the total number of check-ins of location $\loc$.
More popular a location is, higher location entropy it has.

\begin{figure}[!t]
\centering
\begin{minipage}[t]{0.65\columnwidth}
   \centering
   \includegraphics[width=0.95\columnwidth]{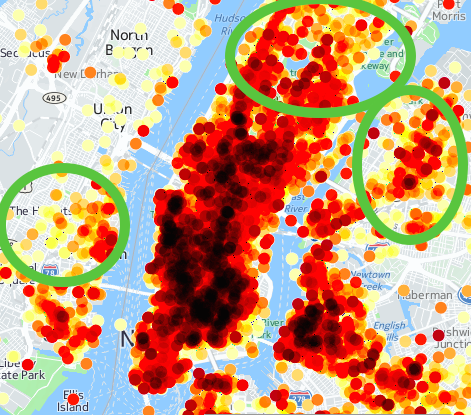}
   \subcaption{Location entropy\label{fig:ny_hmaple}}
\end{minipage}
\begin{minipage}[t]{0.65\columnwidth}
  \centering
  \includegraphics[width=0.95\columnwidth]{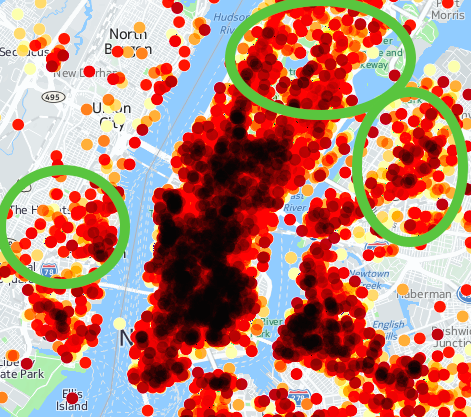}
  \subcaption{Location sociality\label{fig:ny_hmapls}}
\end{minipage}
\caption{Heatmaps in New York.\label{fig:heatmap}}
\end{figure}

To check the difference between the two measurements,
we plot the heatmaps w.r.t.\ location entropy and sociality of New York.
As expected, midtown and downtown New York are ``hot'' areas in both maps.
On the other hand, we observe that
location sociality is more uniformly distributed than location entropy.
For example, the areas marked by green circles in Figure~\ref{fig:ny_hmaple}
are obviously lighter than those in Figure~\ref{fig:ny_hmapls}.
After having a close look, we discover that bars and restaurants are the ``hot'' locations inside these areas.
Data in Los Angeles exhibits a similar result and is not shown.

We further extract the top 20 popular locations in both cities (listed in the additional material).
In New York, the most popular locations are parks
and museums (e.g., The MET, MoMA and Guggenheim).
On the other hand, in Los Angeles, the most popular locations concentrate on malls followed by museums.
Moreover, in both cities, famous landmarks have high location entropy,
such as Rockefeller Center in New York and Hollywood Walk of Fame in Los Angeles.
This is quite different from the ranking in Table~\ref{table:toploc} and Table~\ref{table:topcat}:
social locations are mainly music venues and nightclubs
while popular locations are mainly tourist attractions.
In the end, we conclude that there exists a large difference between social and popular locations.

\section{Case Studies}
\label{sec:casesstudy}

In this section, we perform two case studies to show how location sociality
can be used to build real world applications.
The applications we focus on 
are friendship prediction and new location recommendation,
both of which are essential for social network services.

\subsection{Friendship prediction}
\label{sec:linkpre}
Following the seminal work of Liben-Nowell and Kleinberg,
friendship prediction has been extensively studied~\cite{LK07},
resulting in appealing applications such as friendship recommendation
which is essential for OSNs to increase user engagement.
During the past five years, with the development of LBSNs,
many researchers start to exploit users' location data
as a new source of information for friendship prediction.

\smallskip
\noindent\textbf{Model.}
In our case study, we consider friendship prediction as a binary classification problem.
Each pair of friends is treated positive if they are friends (mutually following each other in Instagram) 
and negative otherwise.
We extract the common locations of two users
and construct the feature space based on these common locations.
Here, two users' common locations
are the intersection of the places they have checked in, regardless of time.
For two users $\usr_i$ and $\usr_j$, 
we find their common locations' sociality
and utilize the average, maximal, minimal and standard deviation
of theses sociality as features for classification.

\begin{figure}[!t]
\centering
\begin{minipage}[!t]{0.85\columnwidth}
  \centering
  \includegraphics[width=1\columnwidth]{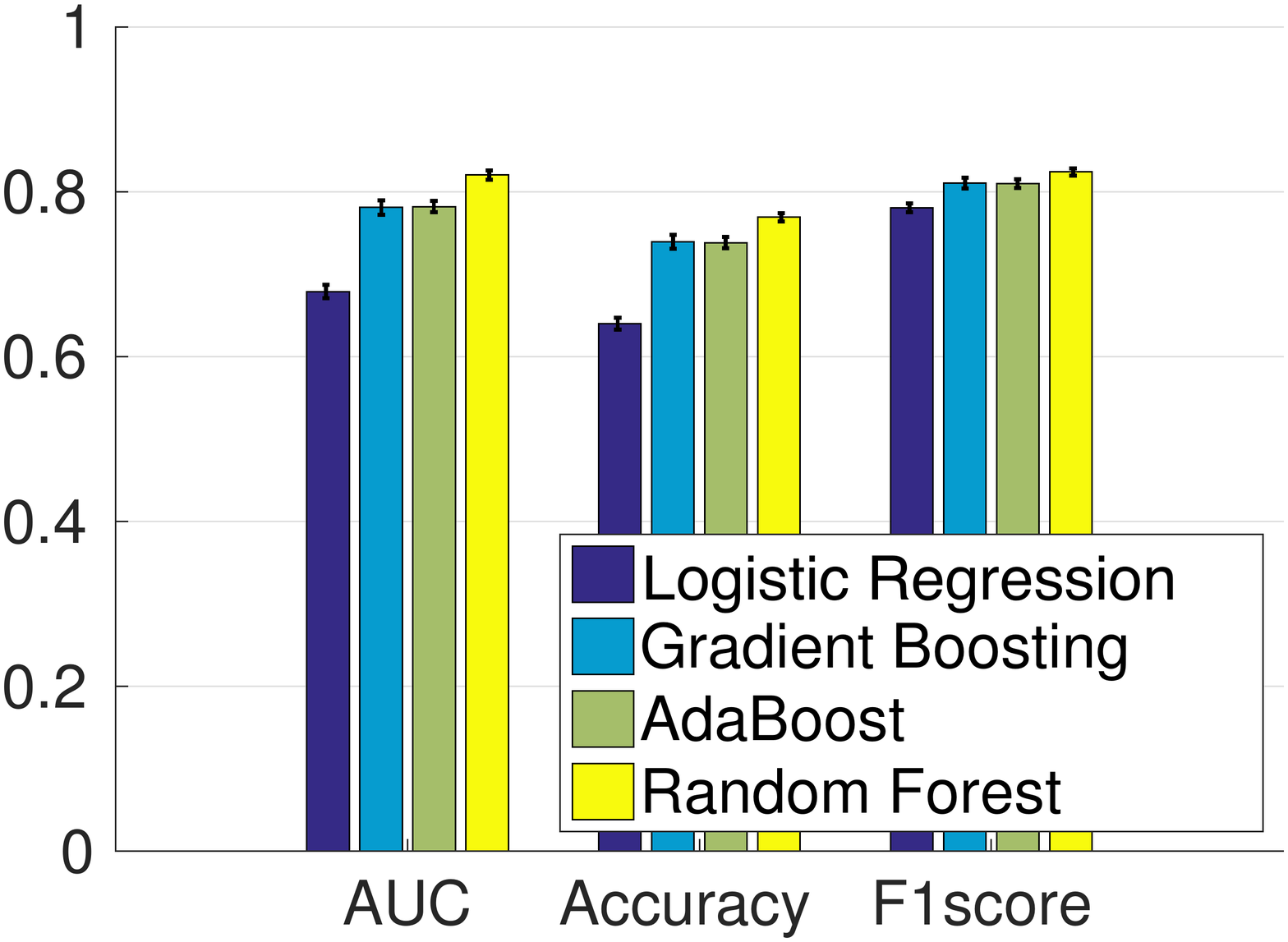}
  \subcaption{New York \label{fig:ny_linkpre_classify}}
\end{minipage}
\begin{minipage}[!t]{0.85\columnwidth}
  \centering
  \includegraphics[width=1\columnwidth]{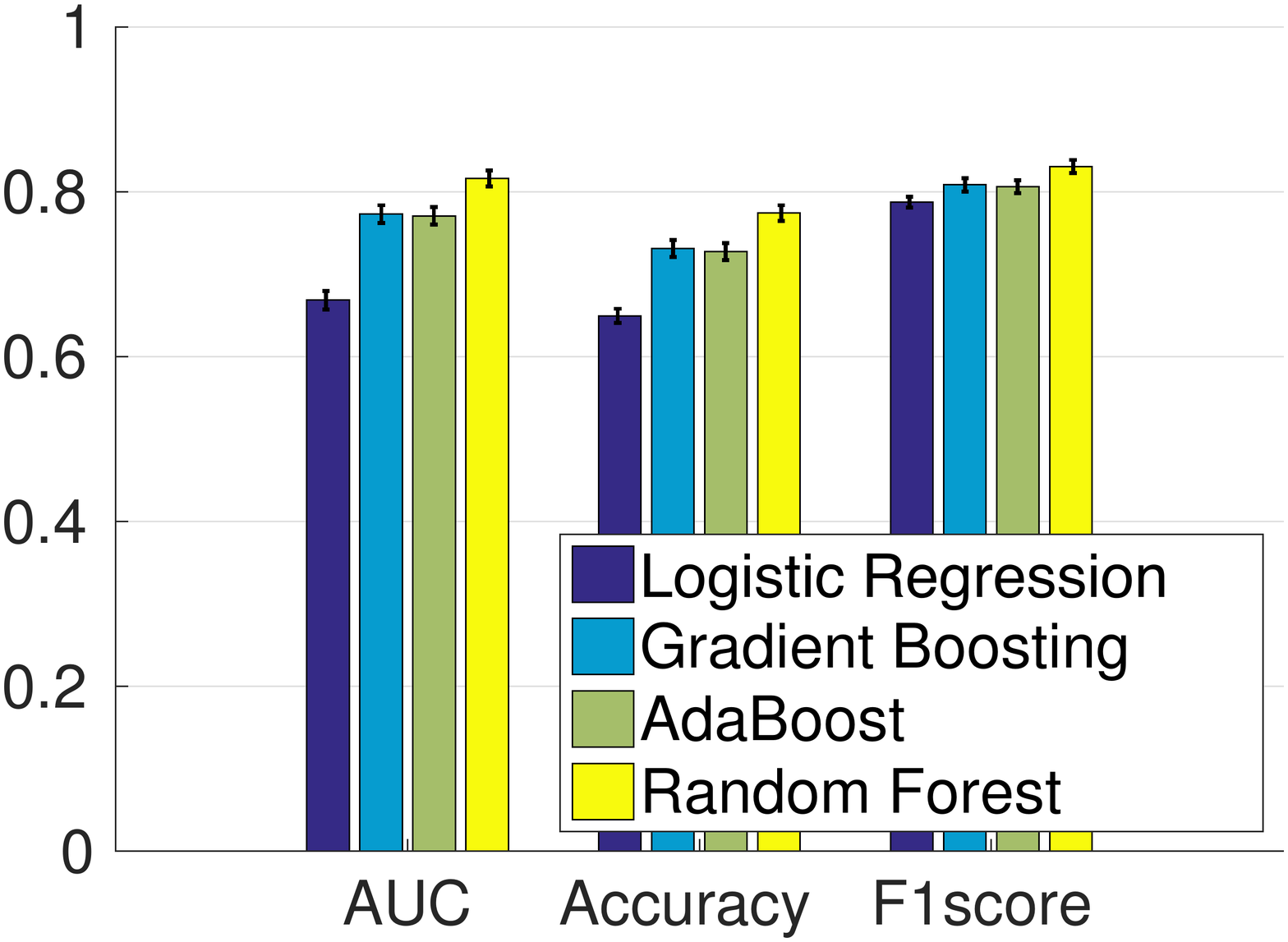}
  \subcaption{Los Angeles\label{fig:la_linkpre_comloc}}
\end{minipage}
 \caption{Evaluation results on our model w.r.t.\ four classification algorithms.\label{fig:linkpre_classify}}
\end{figure} 

\begin{figure}[!t]
\centering
\begin{minipage}[!t]{0.85\columnwidth}
  \centering
  \includegraphics[width=1\columnwidth]{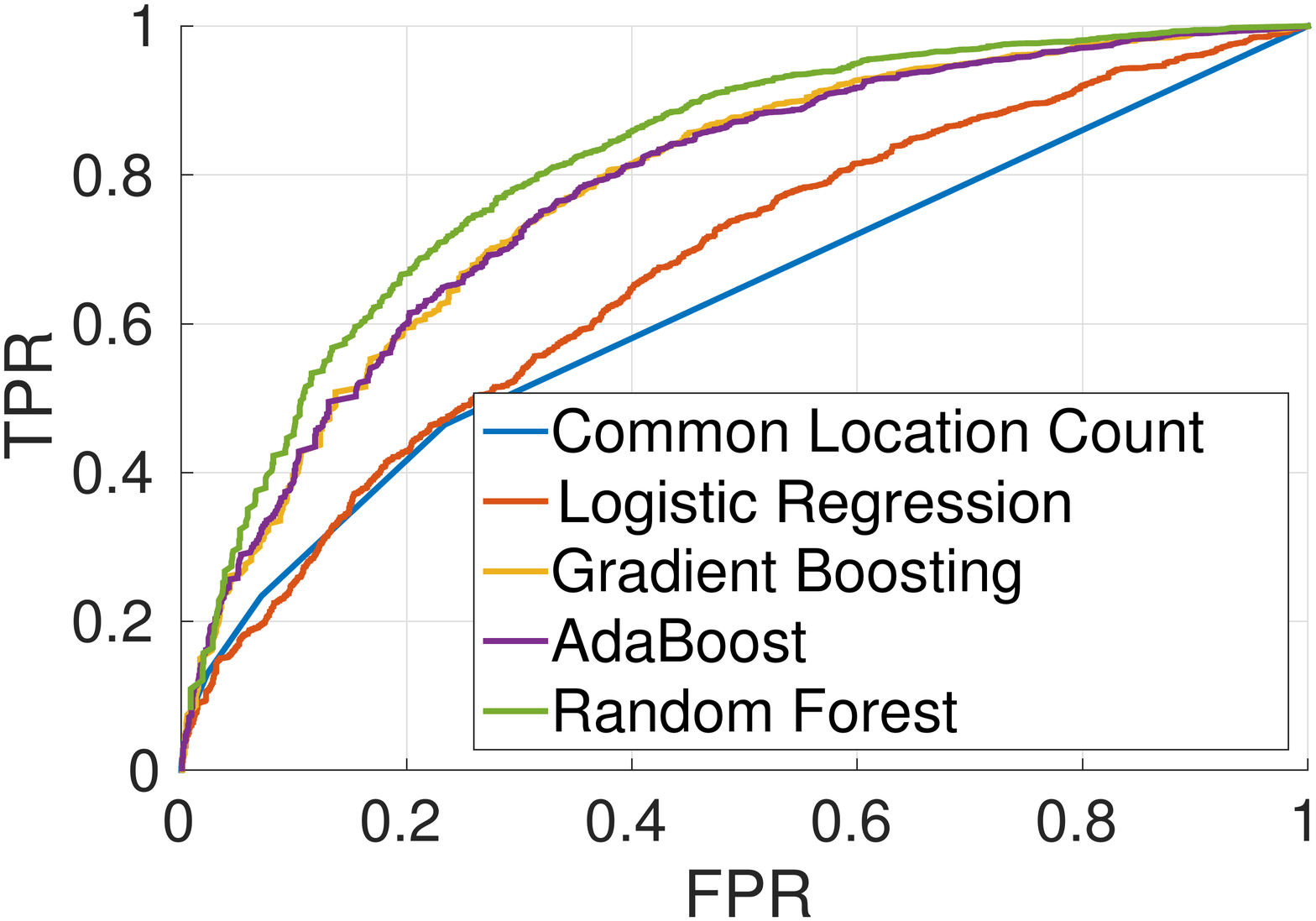}
  \subcaption{New York \label{fig:ny_linkpre_comloc}}
\end{minipage}
\begin{minipage}[!t]{0.85\columnwidth}
  \centering
  \includegraphics[width=1\columnwidth]{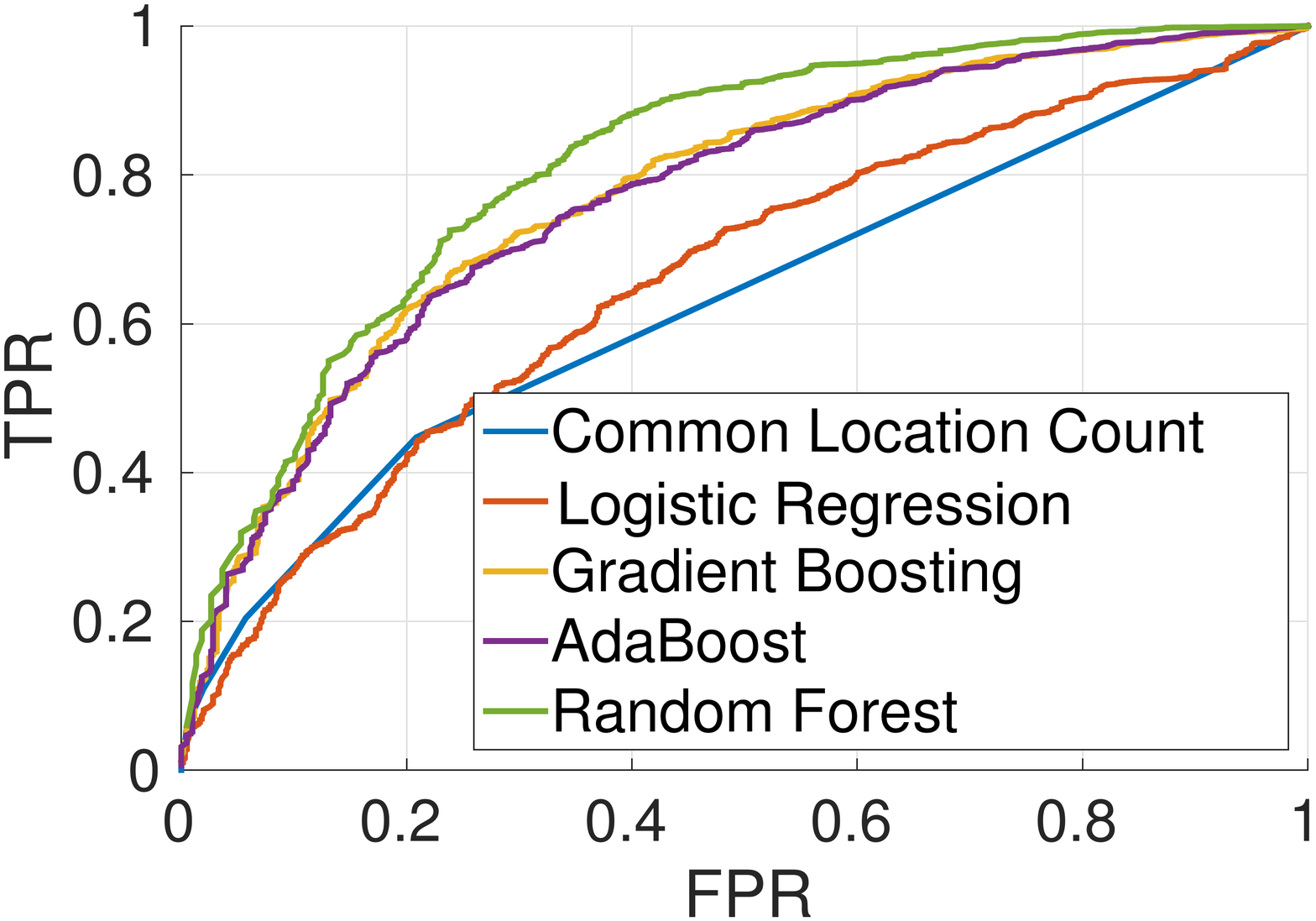}
  \subcaption{Los Angeles\label{fig:la_linkpre_comloc2}}
\end{minipage}
 \caption{ROC curves for our classification and the number of common locations.\label{fig:linkpre_comloc}}
\end{figure} 

\smallskip
\noindent\textbf{Experiment setup.}
To resolve the data sparseness issue, 
we filter out pairs of users who have only one or zero common location.
This leaves us 7,525 pairs of friends, i.e., positive cases, in New York
and 3,961 pairs of friends in Los Angeles.
For negative case, we randomly sample the same number of non-friend pairs --
each pair of them has at least two common places as well.
It is worth noticing that this way of sampling negative cases
increases the hardness of classification
since a non-friend pair also has at least two common locations.
Therefore, we can further evaluate the usefulness of location sociality.
We have adopted four classification algorithms in our experiments
including logistic regression, gradient boosting, AdaBoost and random forest.
Accuracy, F1score and AUC (area under the ROC curve) are used as our metrics.
We randomly split the dataset with 70\% for training and 30\% for testing,
this random split is repeated for 10 times and we report the average result.

\smallskip
\noindent\textbf{Results.}
Figure~\ref{fig:linkpre_classify} depicts the performance of our classifications.
Among all the classifiers, random forest performs the best 
with AUC = 0.82 and Accuracy = 0.77 in the two cities.
Meanwhile, we have F1score = 0.82 in New York and F1score = 0.83 in Los Angeles.
AdaBoost and gradient boosting have a comparable performance.
On the other hand, logistic regression performs the worst.

The number of common locations is further adopted as a naive baseline model for comparison,
i.e., we tune the threshold (the number of common locations) for classification to obtain the ROC curve.
Figure~\ref{fig:linkpre_comloc} plots the results.
As we can see, all our classifiers based on location sociality 
outperform this naive baseline.

Next, we check whether adding location sociality 
into a state-of-the-art model as proposed 
in~\cite{SNM11}
 \footnote{Most of the recent solutions focus on two users' meeting events
 i.e., two users at the same location at roughly the same time,
 while~\cite{SNM11} and our model only exploit common locations.}
can increase prediction performance.
The model in~\cite{SNM11} (location feature setting) 
extracts two users' common locations
and design features mainly with these common locations' entropies (see~Section~\ref{sec:lsvsle} for location entropy), 
such as the minimal location entropy.
In our experiments,
we combine our four location sociality features with the features in~\cite{SNM11}
and fit them into our best performing classifier random forest.
The results in Table~\ref{table:linkpre_masco} 
show that the classification with location sociality improves~\cite{SNM11}
by around 5\% among all three metrics in both New York and Los Angeles.
This further demonstrates that location sociality is useful for friendship prediction.

\begin{table}[!t]
 \centering
 \begin{tabular}{ c | r | r | r }
\toprule[\thickline]
 New York & AUC & Accuracy & F1score\\
\midrule[\thinline]
 \cite{SNM11} & 0.83 & 0.77 & 0.82 \\
 sociality+\cite{SNM11} & \textbf{0.87} & \textbf{0.81} & \textbf{0.85} \\
\bottomrule[\thickline]
\toprule[\thickline]
  Los Angeles & AUC & Accuracy & F1score\\
\midrule[\thinline]
 \cite{SNM11} & 0.82 & 0.77 & 0.82 \\
 sociality+\cite{SNM11} & \textbf{0.86} & \textbf{0.80} & \textbf{0.85} \\
\bottomrule[\thickline]
 \end{tabular}
 \caption{Evaluation result on~\cite{SNM11} and location sociality+\cite{SNM11}~\label{table:linkpre_masco}.}
\end{table}

Indeed, there exist other solutions for friendship prediction
such as considering two users' meeting events~\cite{PSL13,WLL14,HYL15}.
However, the main issue for this method is that meeting events (reflected in OSNs) are rare,
for example, with 6M check-ins in New York, we only observe around 100 meeting events.
Moreover, we want to emphasize that friendship prediction is not the focus of the current work.
Therefore, we choose the most straightforward features for training our classifiers.
Nevertheless, our prediction still achieves a strong performance
showing that location sociality is a good indicator for recommending friends.
Further investigation on integrating location sociality 
into the state-of-the-art friendship prediction models is worth studying
and we leave it as a future work.

\subsection{Location recommendation}
\label{sec:locrecomm}
The second case study we perform 
is recommending new locations for users to visit.
Location recommendation
has a great potential to build appealing applications.
During the past five years, it has attracted academia a lot of attention (e.g.~\cite{ZZXY10,GTHL13,GTHL15,BZWM15}).
Our goal here is to demonstrate the usefulness of location sociality in recommending new locations.
In order to integrate location sociality into a location recommender,
we adopt a classical approach, 
namely random walk with restart~\cite{TFP06}.

\smallskip
\noindent\textbf{Model.}
In a typical setting of random walk with restart for recommendation,
in the beginning
we define a matrix $Q$ as
\[
Q = \left( \begin{array}{cc} 0 & \ulmat \\
\lumat & 0 \end{array} \right)
\]
where $\ulmat$ and $\lumat$ 
represent user-location network (location-user network) (Section~\ref{sec:framework}).
Meanwhile, $\bar Q$ denotes the column stochastic version of $Q$~\cite{YSLYJ11}.
Then to recommend locations to a user $\usr_i$,
we modify $\bar Q$ 
to allow every node in the graph having a certain probability (15\% in the experiments)
to jump to the node representing $\usr_i$.
Formally, for every $Q_{a,b} \in Q$, $\bar Q_{a, b}$ is defined as
\[
 \bar Q_{a, b} = \left\{ \begin{array}{lr}
  (1-c)\cdot \frac{Q_{a, b}}{\sum_j Q_{j, b}} + c \cdot 1& \mbox{~~~~~~~~~if the $a$th row represents } \usr_i \\ 
  (1-c)\cdot\frac{Q_{a, b}}{\sum_j Q_{j, b}} & \mbox{~~~otherwise}
  \end{array}\right.
\]
where $c=0.15$.
By applying the same method of solving PageRank, e.g., power method, 
we can obtain the steady state distribution over $\bar Q$,
which is the relevance score of all nodes (both locations and users) to $\usr_i$.
Locations with high relevance scores are recommended to $\usr_i$.
Noulas et al.~\cite{NSM12} have exploited this approach for 
location recommendation\footnote{They~\cite{NSM12} also consider social network in $Q$,
here we ignore it for better demonstrating location sociality's usefulness.},
where the weight on an edge between a user $\usr_i$ and a location $\loc_j$
is simply the user's number of visits to that location,
i.e., $\ulmat_{i, j} = \w^{\Usr, \Loc}_{i, j}=\cicnt{\usr_i, \loc_j}$ in Section~\ref{sec:framework}.

To integrate location sociality into the edge weight 
for location recommendation,
we change $\ulmat$ to $T$, i.e., $Q$ is modified to:
\[
  Q = \left( \begin{array}{cc} 0 & T \\
  \mathcal{T} & 0 \end{array} \right)
\]
where $T_{i, j}$ is defined as 
\begin{equation} 
\label{equ:locrecomm}
 T_{i, j}  = \cicnt{\usr_i, \loc_j}\cdot\frac{1}{-\log(\ls{\loc_j})}.
\end{equation}
Here, $\ls{\loc_j}$ is the location sociality of $\loc_j$,
meanwhile $\mathcal{T}$ is the transpose of $T$.
Under this formulation, 
Equation~\ref{equ:locrecomm} assigns higher weight to locations with high sociality,
which will bias the recommended locations to be more social.
In the end, by performing power method on the column stochastic version of the modified $Q$,
we obtain the recommended locations for each user.

\smallskip
\noindent\textbf{Experiment setup.}
The check-in dataset is partitioned temporally with each one covers consecutively 60 days~\cite{NSM12}.
For each partition, we use the data of the first 30 days to train the model
while the left 30 days for testing.
Since our aim is to perform new location recommendation,
for each user we further filter out his locations in the testing set 
that he has already been to in the training set.
In the end, we perform random walk with restart with location sociality ({\tt rwr-ls}) 
to recommend locations for each user,
and exploit rwr without location sociality ({\tt rwr}),
i.e., the one in~\cite{NSM12}, as the baseline model.
Two metrics including precision@10 and recall@10 are adopted for evaluation.

\begin{table*}[!t]
 \centering
 \begin{tabular}{c | r | r || c | r | r}
\toprule[\thickline]
 \multicolumn{6}{c}{\textbf{New York}}\\
\midrule[\thinline]
 15.8-15.10 & ~Precision@10~ & ~Recall@10~ & 15.11-16.1 & ~Precision@10~ &~Recall@10~ \\
\midrule[\thinline]
 {\tt rwr} & 0.009 & 0.021 & {\tt rwr} &  0.009 & 0.028 \\
 {\tt rwr-ls} & \textbf{0.010} & \textbf{0.024} & {\tt rwr-ls} & \textbf{0.010} & \textbf{0.031}\\
\midrule[\thinline]
 15.9-2015.11 & Precision@10 & Recall@10 & 15.12-16.2 & Precision@10 & Recall@10 \\
\midrule[\thinline]
 {\tt rwr} & 0.010 & 0.032 & {\tt rwr} &  0.009& 0.026\\
 rwr-ls & \textbf{0.011} & \textbf{0.034} & {\tt rwr-ls} & \textbf{0.010} & \textbf{0.028} \\
\midrule[\thinline]
 15.10-15.12 & Precision@10 & Recall@10 & 16.1-16.3 & Precision@10 &Recall@10 \\
\midrule[\thinline]
 {\tt rwr} & 0.009 & 0.028 & {\tt rwr} &  0.008 & 0.027 \\
 rwr-ls & \textbf{0.010} & \textbf{0.029} & {\tt rwr-ls} & \textbf{0.009} & \textbf{0.029} \\
\bottomrule[\thickline]
\toprule[\thickline]
 \multicolumn{6}{c}{\textbf{Los Angeles}}\\
\midrule[\thinline]
 15.8-15.10 & Precision@10 & Recall@10 & 15.11-16.1 & Precision@10 &Recall@10 \\
\midrule[\thinline]
 {\tt rwr} & 0.010 & 0.028 & {\tt rwr} &  0.008 & 0.024 \\
 rwr-ls & \textbf{0.011} & \textbf{0.030} & {\tt rwr-ls} & \textbf{0.009} & \textbf{0.026}\\
\midrule[\thinline]
 15.9-2015.11 & Precision@10 & Recall@10 & 15.12-16.2 & Precision@10 & Recall@10 \\
\midrule[\thinline]
 {\tt rwr} & 0.013 & 0.038 & {\tt rwr} &  0.012& 0.047\\
 rwr-ls & \textbf{0.015} & \textbf{0.042} & {\tt rwr-ls} & \textbf{0.013} & \textbf{0.048} \\
\midrule[\thinline]
 15.10-15.12 & Precision@10 & Recall@10 & 16.1-16.3 & Precision@10 &Recall@10 \\
\midrule[\thinline]
 {\tt rwr} & 0.010 & 0.025 & {\tt rwr} &  0.007 & 0.035 \\
 rwr-ls & \textbf{0.012} & \textbf{0.029} & {\tt rwr-ls} & \textbf{0.009} & \textbf{0.044} \\
\bottomrule[\thickline]
 \end{tabular}
 \caption{Precision@10 and recall@10 for location recommendation.\label{table:locrecomm}}
\end{table*}

\smallskip
\noindent\textbf{Results.}
Table~\ref{table:locrecomm} presents the results for location recommendation in both cities.
As we can see, {\tt rwr-ls} outperforms {\tt rwr} in all months.
For precision@10, {\tt rwr-ls} outperforms {\tt rwr} by 10\%,
while for recall@10, even in the worst case in New York, 
{\tt rwr-ls} still has 3.4\% improvement on {\tt rwr}.
Even though the absolute precision and recall of our recommendation is not high,
it is worth noticing that the similar performances of location recommendation 
have been obtained by~\cite{YLL12,LX13,GTHL15},
thus our results are reasonable.
Similar to Gao et al.~\cite{GTHL15}, 
we emphasize that the focus here is to compare the relative performance,
in order to demonstrate the usefulness of our quantification.

Many state-of-the-art algorithms exploit other factors for location recommendation
such as geographical distance and users' published contents,
one of our future works is to integrate location sociality into these algorithms
to further improve recommendation.

\section{Implications and Limitations}
\label{sec:implimit}
We discuss implications and limitations of the current work.

\smallskip
\noindent\textbf{Implications.}
Location sociality as a measurement can characterize
a social map for a city.
The applications based on it, including the friendship prediction
and location recommendation addressed in Section~\ref{sec:casesstudy},
can benefit several parties.

For city administrator, 
understanding where people prefer to socialize 
can help them make better city plans.
For example, during conventional social time such as Friday nights,
the city government could make specific transportation plans,
such as more buses and taxis, or security plans such as deploying more policemen,
for areas with high sociality.
Moreover, location sociality can be a good reference
for the government to plan future city development.
For city residents, location sociality provides
a good reference for them to find or discover new places 
to socialize with their friends.
In addition, location sociality can be used 
as an important factor for location recommendation services
for social network services, such as Yelp and Foursquare.
For visitors, visiting high sociality places 
is a good way to engage local people's social life.
This can help visitors understand the city's culture
in a better way.
For business owners, 
knowing where people like to go to conduct social activities
is an important factor for them to determine where to open new business.

\smallskip
\noindent\textbf{Limitations.}
We point out the following limitations of the current work.
First, we only focus on the data from Instagram
which cannot reflect the general population both socially and geographically.
Socially, the authors of~\cite{SCFYCQA15,NZHP16}
have shown that most of Instagram users are center around a younger age (25 yeas old);
geographically, most of the check-ins concentrate on the city center 
(Figure~\ref{fig:nycheckin}).

Second, our quantification does not take into account the temporal factor
which could also be important for understanding locations.
One approach for considering time 
would be concentrating on users' meeting events, 
i.e., users check in at the same location at the same time,
however, as we have discussed previously,
meeting events are rare even in large datasets as ours.
In the future, we plan to incorporate data with rich temporal information
from other sources to further improve our quantification.

Third, we quantify a location's sociality based on its visitors' information.
In some cases, a location's own property can also contribute to its sociality.
For example, it is pointed out that the decoration,
the space allocation and even the bicycle parking design can result in
different number of visitors to a cafe located in a university campus~\cite{M98}.
However, we argue that data of this kind is hard and expensive to obtain at a large scale.

\section{Related Work}
\label{sec:rel}

The emergence of LBSNs has brought us an unprecedented opportunity
to study human mobility and its interaction with social relationships.
Many works have been done on
understanding human mobility and its interaction with social relations.
There mainly exist two research directions:
one is to exploit users' location information
to understand social relations, e.g., see~\cite{SNM11,PSL13,WLL14,HYL15,ZP15}; 
the other is to use social relations to understand mobility and locations,
including the current work.

Backstrom et al.~\cite{BSM10} present one of the pioneer works 
on friendship-based location prediction.
They analyze Facebook users' home location
and discover that friends tend to live closer to each other than strangers.
Then they build a maximal likelihood estimator
to predict a user's home location.
They have shown that their model outperforms significantly the method based on IP addresses.
Cho et al.~\cite{CML11} study a general problem: 
instead of predicting home location, 
they aim to predict where a user is at a certain time.
They construct a dynamic Gaussian mixture model
with the assumption that each user's mobility is centered around two states,
such as home and work.
The experimental results show that 
their model achieves a promising accuracy.
Other works include~\cite{GTL12,J13,PZ15a,PZ15b}.
More recently, Jurgens et al.~\cite{JFMXR15} perform a comprehensive study on most of the existing works
in the field and points out some future directions.

Besides predicting a user's location,
researchers begin to advance our understandings of locations
based on the data from social networks.
In~\cite{QSA14}, the authors focus on recommending pleasant paths between two locations in a city.
Unlike the traditional shortest path recommendation, 
they assign three values to describe whether a street is quiet, beautiful and happy, respectively.
Then they adjust the path recommendation algorithm with these factors
and recommend the most pleasant path for users.
In~\cite{QASD15}, the authors quantify whether a street is suitable for walk,
namely walkability.
To assess their results, they propose to use concurrent validity.
Their discoveries, to mention a few, include walkable streets 
tend to be tagged with walk-related words on Flickr
and can be identified by location types on those streets.
The authors of~\cite{FGM15} exploit the data from Foursquare
to analyze different neighborhoods in a city.
They extract some signature features to profile each neighborhood
and propose an algorithm to match similar neighborhoods across different cities.
Experimental results show that they are able to match tourists areas across Paris and Barcelona,
and expensive residential areas in Washington D.C.\ and New York.
More recently,
Hristova et al.~\cite{HWMPM16} propose four location measurements,
including brokerage, serendipity, entropy and homogeneity,
under a heterogenous social and location network model.
Their experiments are conducted with a Foursquare dataset collected in London,
and the authors show that their proposed measurements
can be used to describe dynamics that is hard to capture 
including gentrification and deprivation.
Other recent works include~\cite{GNM14,QSMM15,HLL15,OHSHH17}.

The current work also falls into the field of urban informatics~\cite{ZCWY14},
a newly emerging field where researchers tend to use the ubiquitous data
to understand and improve the city where we live.
Besides the research literature, 
several open projects have been established as well.
To mention a few examples,
Yuan et al.~\cite{YZX12} focus on discovering the function of each region in a city;
Venerandi et al.~\cite{VQCQS15} measure the socio-economic deprivation of a city.
Another excellent example is the goodcitylife project\footnote{\footnotesize\url{http://goodcitylife.org/}},
where the team members try to 
imitate human beings' five senses on food 
to understand cities.

\section{Conclusion}
\label{sec:conclu}
In this paper, we have proposed a new notion namely location sociality
to describe whether a location is suitable for conducting social activities.
We constructed a heterogenous network linking locations and users 
and proposed a mixture model of HITS and PageRank to quantify location sociality.
Experimental results on millions of Instagram check-in data validate location sociality with some in-depth discoveries.
Two case studies including friendship prediction and location recommendation
demonstrate the usefulness of our quantification.

Location data do not only come from LBSNs,
but many other sources, such as GPS traces and WIFI points.
In the future, 
we are  interested in establishing more connections between LBSN data and other sources
to gain a deep understanding of cities.

\bibliographystyle{ACM-Reference-Format}
\bibliography{ht2017} 

\end{document}